\documentclass[12pt]{article}
\usepackage{graphics}
\usepackage{fancyheadings}

\begin{document}

\title{A description of several coordinate systems for hyperbolic spaces}
\author{Sandro S. e Costa \\
{\small {\it Instituto Astron\^omico e Geof\'\i sico da Universidade de
S\~ao Paulo} (IAG-USP)}\\
{\small {\it Av. Miguel St\'efano, 4200 - CEP 04301-904 - S\~ao Paulo - SP -
Brazil}}\\
{\small {\it Electronic address:} {\tt sancosta@iagusp.usp.br}}}
\date{\today }
\maketitle

\begin{abstract}
This article simply presents several coordinate systems for 2 and
3-dimensional hyperbolic spaces, describing the general solutions of
Helmholtz equation in each one of these systems.

$\,\,$

\centerline{PACS numbers:  02.40.-k}
\end{abstract}

\section{Introduction}

Although the most recent cosmological observations suggest we live in a flat
universe, the possibility that such observations are indicating a non-flat
space with almost vanishing curvature can not yet be excluded, since it is
not easy to distinguish a truly flat universe, where the curvature is null,
from a negatively or positively curved universe inflated until the point of
having almost no curvature.

Such indeterminacy, somewhat fundamental, is linked to another one about the
compactness or not of the universe, {\it i.e.}, of its topology, and in any
case one can ask if it would be possible, in principle, to achieve a higher
degree of certainty.

The line of research confronting such problem can be summarized by the
question ``can one see the shape of the universe?'' \cite{Lachieze}, which
is related to the problem of ``hearing the shape of a drum'' \cite{Kac,GY}.
Unfortunately, the answers obtained until now from the cosmological
observations are unconclusive or, at most, not much restrictive. More
complete studies would, for example, involve as a first step the solution of
the eigenvalue problem of the Laplace-Beltrami operator, {\it i.e.}, the
Helmholtz equation, in several spaces of different curvatures, compact and
not compact, and this is yet an open field of research \cite{Inoue,Cornish2}.

For flat, Euclidean geometry, it is easy to find in textbooks all the
systems of coordinates where the Helmholtz equation is solvable\footnote{
In the case of Euclidean space the interested reader can find a good
description of how to obtain all the coordinate systems in which the
Helmholtz equation is separable, involving the use of complex variables and
linear algebra, in the textbook of Morse and Feshbach \cite{Morse};
see also the work of Eisenhart \cite{Eisenhart}.
Here, we simply {\it describe some} coordinate systems for
hyperbolic spaces.}: 4 systems
for the 2-dimensional Euclidean space $E^2$ and 11 systems for the
3-dimensional Euclidean space $E^3$ \cite{Morse,Arfken}. In contrast to this, one
can see that almost all physically motivated studies of hyperbolic spaces
present in literature use only the analogous of the spherical polar
coordinates of flat geometry, probably because such coordinate system is
obtained as a trivial generalization for hyperbolic geometry, and also
probably because this system of coordinates is the one one would expect to
use, for instance, in cosmological observations.

However, in the same way as in flat geometry, there are also other groups of
coordinates in hyperbolic geometry which can be adequate for several
specific problems, or where the Helmholtz equation can be solved. One recent
example found in literature is the use of a set of hyperbolic coordinates
analogous to the Euclidean cylindrical ones to solve a problem of quantum
cosmology \cite{Costa}.

With no other purpose than serving as one more tool for works involving 2
and 3-dimensional hyperbolic spaces, and with no intention of being
complete, this paper simply presents a description of several sets of
coordinates for these spaces, showing in the majority of cases the general
solutions of Helmholtz equation for the sets presented -- 4 sets for the
2-dimensional space $H^2$ and 12 for the 3-dimensional space $H^3$, numbers
which probably do not cover all possibilities for the spaces $H^2$ and $H^3$.

The paper is organized as follows: section \ref{pro} describes two types of
projections that can be used for visualizing the hyperbolic space; sections 
\ref{2d} and \ref{3d} present, respectively, different coordinate systems
for 2 and 3-dimensional hyperbolic spaces; and section \ref{end} is a brief
conclusion.

\section{\label{pro} Projections of the hyperbolic space}

The infinite hyperbolic space $H^n$ can be described by the $n$-dimensional
surface described by the constraint 
\begin{equation}
\label{vinculo}x^\mu x_\mu \equiv g_{\mu \nu }x^\mu x^\nu =\left( x^1\right)
^2+\left( x^2\right) ^2+...-\left( x^{n+1}\right) ^2=-1\;\;, 
\end{equation}
where the last equality shows explicitly that $g_{\mu \nu }=\;$diag$
\,\left[ 1,...,1,-1\right] $ is the metric of a Minkovski space of dimension $%
n+1$, also represented by the element of line 
\begin{equation}
ds^2=\left( dx^1\right) ^2+...+\left( dx^n\right) ^2-\left( dx^{n+1}\right)
^2\;\;. 
\end{equation}
Notice that equation (\ref{vinculo}) is in perfect analogy with the
constraint equation for a spherical surface, $x^\mu x_\mu =1$, where the
metric used is the Euclidean one.

In order to better visualize the infinite hyperbolic surface one can use
several types of geometric projections, in almost the same way one does to
visualize the spherical surface of the Earth in one flat sheet of paper. Two
kinds of projections, known as Klein and Poincar\'e projections, are such
that the entire $n$-dimensional hyperbolic space is `confined' into a
circular unitary `ball' of dimension $n$. The coordinates for these
projections are defined, respectively, as 
\begin{equation}
x^{\mu }_{K}\equiv \frac{x^{\mu }}{x^{n+1}} 
\end{equation}
and 
\begin{equation}
x^{\mu }_{P}\equiv \frac{x^{\mu }}{1+x^{n+1}}\;\;, 
\end{equation}
where $\mu =1,2,...,n$. In the Klein projection the geodesics of the
hyperbolic space -- which are hyperbolas -- appear as straight lines, while
in the Poincar\'e projection they appear as arcs of circles {\it or}
straight lines.

An artistic example of the use of these projections can be seen
in some works of the Dutch painter M.C. Escher, who used them to present
fillings of the entire infinite hyperbolic plane $H^2$ with repetitive
patterns, a procedure analogous to the covering of a wall with regular tiles\footnote{%
Similar examples are some works by Peter Raedschelders which appear in the
book {\it `Surfing through the hyperspace'}, by Clifford A. Pickover \cite
{Pickover}.}.

In the next section both the Klein and Poincar\'e projections for a
2-dimensional hyperbolic space will be used to show for each system of
coordinates a representation of the curves produced when one of the
coordinates is kept constant.

\section{\label{2d} 2-dimensional hyperbolic space}

\subsection{Symmetries and coordinates}

As said in the previous section, the 2-dimensional hyperbolic space can be
seen as formed by all points that satisfy the constraint 
\begin{equation}
\label{vinculo2}\left( x^1\right) ^2+\left( x^2\right) ^2-\left( x^3\right)
^2=-1\;\;. 
\end{equation}
Notice that the `central point' $\left( 0,0,1\right) $ obeys this
constraint. Notice also that any change of coordinates that keeps this
equation still valid are allowed, producing then parametric equations which
will define a coordinate system.

In fact, any coordinate transformation that keeps equation (\ref{vinculo2})
invariant is a representation of the internal symmetries of hyperbolic space 
$H^2$. One such symmetry is represented by Lorentz boosts in the $x^1$ and $%
x^2$ directions, and other is given by rotations in the plane $x^1x^2$. This
symmetries are represented by the matrices 
\begin{equation}
R\left(\varphi\right) \equiv \left( 
\begin{array}{ccc}
\cos \varphi & \sin \varphi & 0 \\ 
-\sin \varphi & \cos \varphi & 0 \\ 
0 & 0 & 1 
\end{array}
\right) 
\end{equation}
and 
\begin{equation}
\begin{array}{ccc}
\Lambda\left(\varphi ,a\right) = R\left(\varphi \right)\times \left( 
\begin{array}{ccc}
\cosh a & 0 & \sinh a \\ 
0 & 1 & 0 \\ 
\sinh a & 0 & \cosh a 
\end{array}
\right)\times R^{-1}\left(\varphi \right) &  &  
\end{array}
\;\;, 
\end{equation}
which represent, respectively, a rotation of angle $\varphi $ in the plane
and a boost of rapidity $a$ in the direction given by the angle $\varphi $.

One way of combining all these informations at once is to apply, in
sequence, two different orthogonal boosts and a rotation to the point $%
\left( 0,0,1\right)$, obtaining then a general point $P$, 
\begin{eqnarray}
P&=&R\left(\varphi\right)\times\Lambda\left( \frac{\pi }{2},a\right)\times\Lambda\left( 0,b\right)\times
\left(
\begin{array}{c}
0\\ 0\\ 1
\end{array}
\right)\nonumber \\
&=&
\left(
\begin{array}{c}
\cos\varphi\sinh b -\sin\varphi\sinh a\cosh b \\
\sin\varphi\sinh b +\cos\varphi\sinh a\cosh b \\
\cosh a\cosh b
\end{array}
\right)\;\;.
\label{parametrizacoes}
\end{eqnarray}
If $a =0$ or $b =0$ this equation gives the polar parametrization of the
space $H^2$, 
\begin{equation}
\left\{ 
\begin{array}{lcl}
x^1 & = & \sinh\chi\cos\varphi \\ 
x^2 & = & \sinh\chi\sin\varphi \\ 
x^3 & = & \cosh\chi 
\end{array}
\right.\;\;. 
\end{equation}
The element of line obtained for these coordinates is\footnote{%
Here one can use the substitution $\sinh\chi =\tan\theta$, where $-\pi/2 <
\theta < \pi/2 $ is the angle known as Gudermannian or hyperbolic amplitude 
\cite{GR}, to obtain the element of line $ds^2=\sec ^2\theta\left( d\theta
^2+\sin ^2\theta d\varphi ^2\right) $, an equality which shows that these
coordinates produce a metric conformal to the one of the sphere.} 
\begin{equation}
ds^2\equiv\left(dx^1\right)^2+\left(dx^2\right)^2-\left(dx^3\right)^2=d\chi
^2+\sinh ^2\chi d\varphi ^2 \;\;. 
\end{equation}

Putting $\varphi =0$ in (\ref{parametrizacoes}) one obtains another
parametrization, 
\begin{equation}
\left\{ 
\begin{array}{lcl}
x^1 & = & \sinh\rho \\ 
x^2 & = & \cosh\rho\sinh\omega \\ 
x^3 & = & \cosh\rho\cosh\omega 
\end{array}
\right.\;\;, 
\end{equation}
with the element of line\footnote{%
Here the use of the substitution $\sinh\rho=\tan\theta $, where $-\pi/2 <
\theta < \pi/2 $ is again the Gudermannian, gives the element of line $%
ds^2=\sec ^2\theta\left( d\theta ^2+d\omega ^2\right) $.} 
\begin{equation}
ds^2=d\rho ^2+\cosh ^2\rho d\omega ^2 \;\;. 
\end{equation}

Another distinct parametrization of the space $H^2$ can be obtained from the
central point $\left( 0,0,1\right) $ by the product 
\begin{equation}
\left( 
\begin{array}{ccc}
1 & -\mu & \mu \\ 
\mu & 1-\frac{\mu ^2}{2} & \frac{\mu ^2}{2} \\ \mu & -\frac{\mu ^2}{2} & 1+%
\frac{\mu ^2}{2} 
\end{array}
\right) \left( 
\begin{array}{ccc}
1 & 0 & 0 \\ 
0 & \cosh\sigma & \sinh\sigma \\ 
0 & \sinh\sigma & \cosh\sigma 
\end{array}
\right) \left( 
\begin{array}{l}
0 \\ 
0 \\ 
1 
\end{array}
\right)\;, 
\end{equation}
revealing then that the matrix 
\begin{equation}
M\left(\mu \right)\equiv \left( 
\begin{array}{ccc}
1 & -\mu & \mu \\ 
\mu & 1-\frac{\mu ^2}{2} & \frac{\mu ^2}{2} \\ \mu & -\frac{\mu ^2}{2} & 1+%
\frac{\mu ^2}{2} 
\end{array}
\right)\;, 
\end{equation}
with the property $M\left(\mu \right) M\left(\nu \right) =M\left(\mu
+\nu\right)$, inverse $M^{-1}\left(\mu\right) =M\left(-\mu\right) $, and
with $\det\left[M\right] =1$, also represents a symmetry of $H^2$.

The parametrization thus obtained, 
\begin{equation}
\left\{ 
\begin{array}{l}
x^1=e^{-\sigma }\mu \\ 
x^2=\sinh \sigma +e^{-\sigma } 
\frac{\mu ^2}2 \\ x^3=\cosh \sigma +e^{-\sigma }\frac{\mu ^2}2 
\end{array}
\right. \;\;, 
\end{equation}
gives the element of line\footnote{%
It is also interesting to use, in this parametrization, the substitution $%
z=e^{\sigma }$, with $0< z\leq\infty $, what gives $ds^2=\left( d\mu
^2+dz^2\right) /z^2$. Such element of line represents the upper half-space
model of the space $H^2$ \cite{Thurston}.} 
\begin{equation}
ds^2=d\sigma ^2+e^{-2\sigma }d\mu ^2\;\;. 
\end{equation}
Such parametrization appears interestingly related to an aplication in
microwave engineering \cite{Terras}, producing a graph known as `Smith
chart'.

The three parametrizations shown until now are not symmetric in the
coordinates $x^1$ and $x^2$. A fourth parametrization with complete
equivalence between these coordinates is 
\begin{equation}
\left\{ 
\begin{array}{l}
x^1= 
\sqrt{2}\cosh u\sinh v \\ x^2= 
\sqrt{2}\cosh v\sinh u \\ x^3=\sqrt{\cosh 2u\cosh 2v} 
\end{array}
\right. \;\;, 
\end{equation}
with element of line 
\begin{equation}
ds^2=\left( \cosh 2u+\cosh 2v\right) \left( \frac{du^2}{\cosh 2u}+\frac{dv^2%
}{\cosh 2v}\right) \;\;. 
\end{equation}

\begin{figure}[p]
\centerline{\scalebox{.7}{\includegraphics{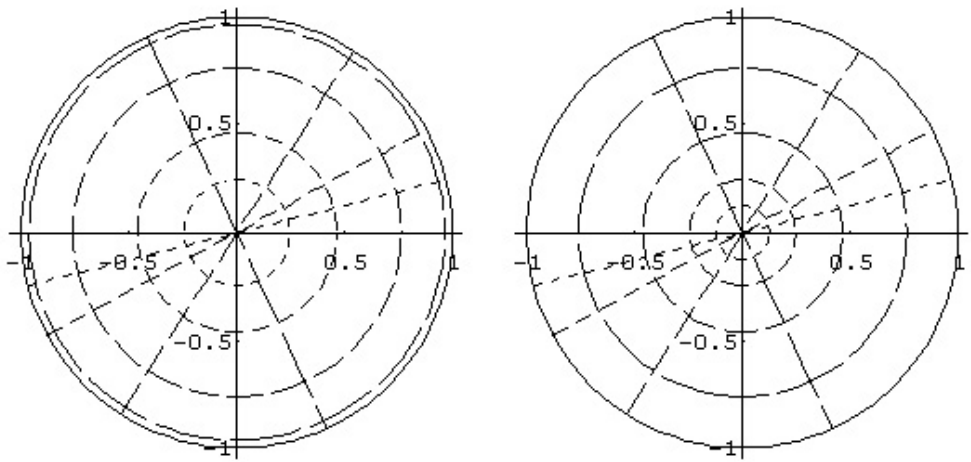}}}
\caption{\label{kp1} Visualization in the projections of Klein (left) and Poincar\'e (right) of the radial coordinates $\left(\chi ,\varphi \right)$: constant $\chi $ produces circles while constant $\varphi $ produces straight lines.}
\end{figure}

\begin{figure}[p]
\centerline{\scalebox{.7}{\includegraphics{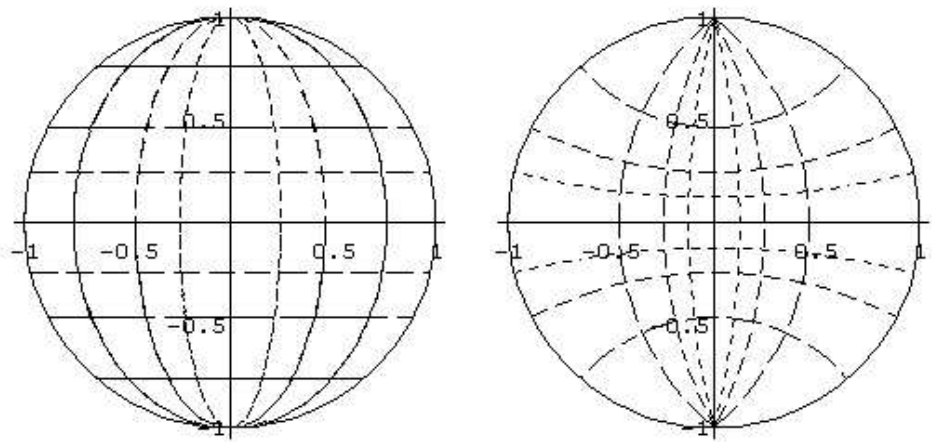}}}
\caption{\label{kp2} Visualization in the projections of Klein (left) and Poincar\'e (right) of the hyperbolic coordinates $\left(\rho ,\omega \right)$.}
\end{figure}

\begin{figure}[p]
\centerline{\scalebox{.7}{\includegraphics{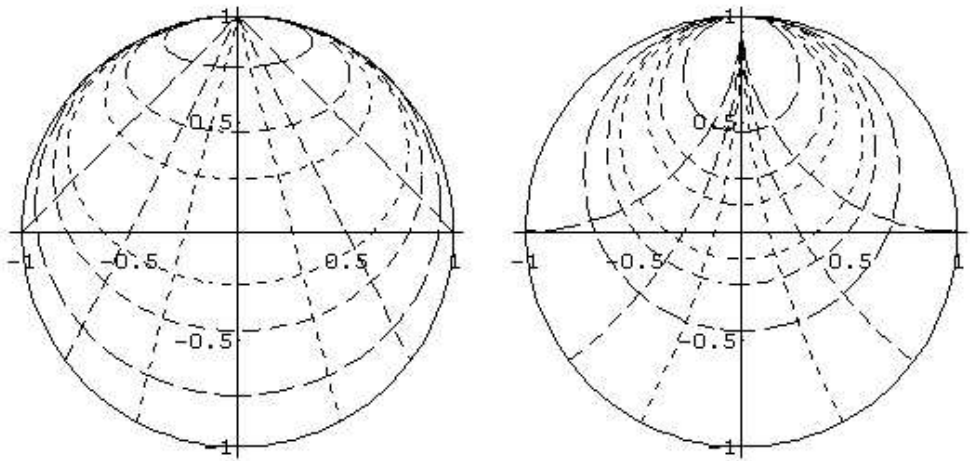}}}
\protect\caption{\label{kp3} Visualization in the projections of Klein (left) and Poincar\'e (right) of the coordinates $\left(\sigma ,\mu \right)$: constant $\sigma $ produces ellipses or circles while constant $\mu $ produces convergent lines. The Poincar\'e projection of these coordinates serves as basis for a graph called ``Smith Chart'', of interest in microwave engineering \protect\cite{Terras}.} 
\end{figure}

\begin{figure}[p]
\centerline{\scalebox{.7}{\includegraphics{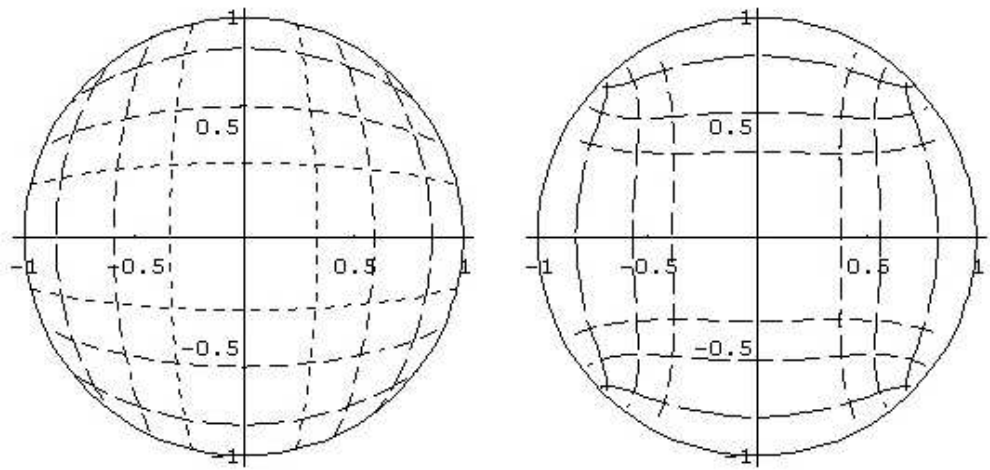}}}
\caption{\label{kp4} Visualization in the projections of Klein (left) and Poincar\'e (right) of the hyperbolic symmetric coordinates $\left( u,v \right)$.}
\end{figure}

Table \ref{t1} presents a summary of the coordinates shown here for the
space $H^2$, with arbitrary names given for each system, while Figures \ref
{kp1} to \ref{kp4} show, in Klein and Poincar\'e projections, the lines
obtained for each set of coordinates, when one of the coordinates is kept
constant.

\begin{table}[t]
\begin{center}
\begin{tabular}{|l|c|l|}
\hline
\multicolumn{3}{|c|}{{\bf Coordinate systems for the hyperbolic 2-D space}} \\ \hline
{\it system} & {\it coords.} & {\it element of line} \\ \hline 
polar & $ \chi ,\varphi $ & $ d\chi ^2+\sinh ^2\chi d\varphi ^2 $ \\ \hline
hyperbolic & $ \rho ,\omega $ & $ d\rho ^2+\cosh\rho ^2d\omega ^2 $ \\ \hline
exponential & $ \sigma ,\mu $ & $ d\sigma ^2 +e^{-2\sigma }d\mu ^2 $ \\ \hline
symmetric & $ u,v $ & $ \left(\cosh 2u +\cosh 2v\right)\left(du^2/\cosh 2u+dv^2/\cosh 2v\right) $ \\ \hline
\end{tabular}
\end{center}
\caption{\label{t1} Summary of systems for space $H^2$.}
\end{table}

\subsection{Helmholtz equation}

The Helmholtz equation \cite{Frankel} 
\begin{equation}
\label{Helmholtz}\nabla ^2\Psi +k^2\Psi \equiv \frac 1{\sqrt{g}}\partial
_\mu \left[ \sqrt{g}g^{\mu \nu }\partial _\nu \right]\Psi +k^2\Psi =0 \;\;, 
\end{equation}
is separable in all four coordinate systems presented here for the space $%
H^2 $.

In polar coordinates $\left( \chi ,\varphi \right) $ one has 
\begin{equation}
\label{c0}\frac 1{\sinh \chi }\left[ \frac \partial {\partial \chi }\left(
\sinh \chi \frac{\partial \Psi }{\partial \chi }\right) +\frac \partial
{\partial \varphi }\left( \frac 1{\sinh \chi }\frac{\partial \Psi }{\partial
\varphi }\right) \right] =-k^2\Psi \;, 
\end{equation}
with solutions of the type $\Psi \left( \chi ,\varphi \right) =X\left( \chi
\right) \Phi \left( \varphi \right) $, with 
\begin{equation}
\label{vinteedoisa}\Phi =a_1\cos \lambda \varphi +a_2\sin \lambda \varphi 
\end{equation}
and 
\begin{equation}
\label{vinteedoisb}X=
b_1P_{-\frac 12+ 
\sqrt{\frac 14-k^2}}^\lambda \left( \cosh \chi \right) +b_2Q_{-\frac 12+%
\sqrt{\frac 14-k^2}}^\lambda \left( \cosh \chi \right) \;\;, 
\end{equation}
or
\begin{equation}
\label{vinteedoisc}X=
\left( \cosh \chi \right) ^{^{-\frac 12}}\left[ b_1^{\prime }P_{-\frac
12+\lambda }^{\sqrt{\frac 14-k^2}}\left( \tanh \chi \right) +b_2^{\prime
}Q_{-\frac 12+\lambda }^{\sqrt{\frac 14-k^2}}\left( \tanh \chi \right)
\right] \;, 
\end{equation}
where $P_\nu ^\mu \left( z\right) $ and $Q_\nu ^\mu \left( z\right) $ are
associated Legendre functions, and where $\lambda $ is just a separation
constant.

For the hyperbolic coordinates $\left( \rho ,\omega \right) $ one has 
\begin{equation}
\label{wrau}\frac 1{\cosh \rho }\left[ \frac \partial {\partial \rho }\left(
\cosh \rho \frac{\partial \Psi }{\partial \rho }\right) +\frac \partial
{\partial \omega }\left( \frac 1{\cosh \rho }\frac{\partial \Psi }{\partial
\omega }\right) \right] =-k^2\Psi \;, 
\end{equation}
an equation solved by the function $\Psi \left( \rho ,\omega \right)
=R\left( \rho \right) W\left( \omega \right) $, with 
\begin{equation}
\label{w}W=a_3\cos \lambda \omega +a_4\sin \lambda \omega 
\end{equation}
and 
\begin{equation}
\label{rau}R=
\left( \cosh \rho \right) ^{-\frac 12}\left[ b_3P_{-\frac 12+i\lambda }^{
\sqrt{\frac 14-k^2}}\left( \tanh \rho \right) +b_4Q_{-\frac 12+i\lambda }^{%
\sqrt{\frac 14-k^2}}\left( \tanh \rho \right) \right] \;\;,
\end{equation}
or
\begin{equation}
\label{rau2}R= 
b_3^{\prime }P_{-\frac 12+\sqrt{\frac 14-k^2}}^{i\lambda }\left( i\sinh \rho
\right) +b_4^{\prime }Q_{-\frac 12+\sqrt{\frac 14-k^2}}^{i\lambda }\left(
i\sinh \rho \right) \;\;,
\end{equation}
with $\lambda $ being again a separation constant. For completeness, it is
important to say that one can also find solutions for $R\left( \rho \right) $
in terms of Gegenbauer or hypergeometric functions by use of the
substitution $\zeta ^{\prime }\equiv \tanh ^2\rho $.

The exponential coordinate system, $\left( \sigma ,\mu \right) $, gives the
equation 
\begin{equation}
\label{vinteeseis}\frac{\partial ^2\Psi }{\partial \sigma ^2}-\frac{\partial
\Psi }{\partial \sigma }+e^{2\sigma }\frac{\partial ^2\Psi }{\partial \mu ^2}%
=-k^2\Psi \;\;. 
\end{equation}
In this case the solutions are $\Psi \left( \sigma ,\mu \right) =\Sigma
\left( \sigma \right) \Theta \left( \mu \right) $, where 
\begin{equation}
\label{vinteesete}\left\{ 
\begin{array}{l}
\Theta \left( \mu \right) =a_{+}e^{i\lambda \mu }+a_{-}e^{i\lambda \mu } \\ 
\Sigma \left( \sigma \right) =e^{\sigma /2}\left[ bI_{\sqrt{\frac 14-k^2}%
}\left( \lambda e^\sigma \right) +cK_{\sqrt{\frac 14-k^2}}\left( \lambda
e^\sigma \right) \right] 
\end{array}
\right. \;\;, 
\end{equation}
with $I_\nu $ and $K_\nu $ being modified Bessel functions (Bessel functions
of imaginary argument). $\lambda $ is again a separation constant.

Finally, the symmetric set of coordinates $\left( u,v\right) $ produces the
equation 
\begin{eqnarray}
&&\cosh 2u\frac{\partial ^2\Psi }{\partial u^2}+\sinh 2u\frac{\partial\Psi }{\partial u }
+\cosh 2v\frac{\partial ^2\Psi }{\partial v^2}+\sinh 2v\frac{\partial\Psi }{\partial v }\nonumber\\
&&+k^2\left(\cosh 2u +\cosh 2v\right)\Psi =0\;,
\end{eqnarray}
which the function $\Psi \left( u,v\right) =U\left( u\right) V\left(
v\right) $ divides into 
\begin{equation}
\label{vinteenove}\frac{d^2U}{du^2}+\tanh 2u\frac{dU}{du}+\left( k^2-\frac{%
\lambda ^2}{\cosh 2u}\right) U=0 
\end{equation}
and 
\begin{equation}
\label{trinta}\frac{d^2V}{dv^2}+\tanh 2v\frac{dV}{dv}+\left( k^2+\frac{%
\lambda ^2}{\cosh 2v}\right) V=0\;. 
\end{equation}
Unfortunately, it is not easy to find simple analytical solutions for these
two equations, except for the case $\lambda =0$, when the solutions are 
\begin{equation}
U\left( u\right) =\left( \cosh 2u\right) ^{-1/4}\left[ aP_{-\frac 14}^{\frac
12 
\sqrt{\frac 14-k^2}}\left( \tanh 2u\right) +bQ_{-\frac 14}^{\frac 12\sqrt{%
\frac 14-k^2}}\left( \tanh 2u\right) \right] 
\end{equation}
and
\begin{equation}
V\left(
v\right) =\left( \cosh 2v\right) ^{-1/4}\left[ a^{\prime }P_{-\frac
14}^{\frac 12\sqrt{\frac 14-k^2}}\left( \tanh 2v\right) +b^{\prime
}Q_{-\frac 14}^{\frac 12\sqrt{\frac 14-k^2}}\left( \tanh 2v\right) \right]
\;\;. 
\end{equation}

\section{\label{3d} 3-dimensional hyperbolic space}

\subsection{Coordinate systems}

The 3-dimensional hyperbolic space can be seen as formed by the set of
points following the 4-dimensional version of the constraint (\ref{vinculo2}%
), 
\begin{equation}
\left( x^1\right) ^2+\left( x^2\right) ^2+\left( x^3\right) ^2-\left(
x^4\right) ^2=-1\;\;. 
\end{equation}

As discussed in the introduction, for the flat Euclidean 3-dimensional space $E^3$
there are eleven coordinate systems in which the Helmholtz equation is
separable \cite{Arfken}. In this section almost the same number of systems
is presented for the hyperbolic $H^3$ space. The presentation follows a
classification based on generalizations of the four basic types of
coordinates found for the 2-dimensional hyperbolic space $H^2$.

\subsubsection{Polar coordinates}

The most known coordinate system for the $H^3$ space is the hyperbolic
equivalent of a sphere, 
\begin{equation}
\label{trintaedois}\left\{ 
\begin{array}{lcl}
x^1 & = & \sinh \chi \cos \varphi \sin \theta \\ 
x^2 & = & \sinh \chi \sin \varphi \sin \theta \\ 
x^3 & = & \sinh \chi \cos \theta \\ 
x^4 & = & \cosh \chi 
\end{array}
\right. \;\;, 
\end{equation}
which produces the element of line 
\begin{equation}
\label{trintaetres}ds^2=d\chi ^2+\sinh ^2\chi \left( d\theta ^2+\sin
^2\theta d\varphi ^2\right) \;\;. 
\end{equation}
Notice that this element of line has a part conformal to the element of line
of a spherical surface.

However, a positively curved surface allows also a parametrization distinct
from the one of the sphere. Such parametrization, of a semi-sphere, is
totally symmetric: 
\begin{equation}
x^1=\sqrt{2}\cos \theta \sin \phi ,\;\;x^2=\sqrt{2}\sin \theta \cos \phi
,\;\;x^3=\sqrt{\cos 2\theta \cos 2\phi }, 
\end{equation}
where $-\pi /4\leq \theta ,\phi \leq \pi /4$. So, changing the spherical
part of the polar hyperbolic coordinates by this one one has 
\begin{equation}
\left\{ 
\begin{array}{lcl}
x^1 & = & \sqrt{2}\sinh \chi \cos \theta \sin \phi \\ x^2 & = & \sqrt{2}%
\sinh \chi \sin \theta \cos \phi \\ x^3 & = & \sinh \chi 
\sqrt{\cos 2\theta \cos 2\phi } \\ x^4 & = & \cosh \chi 
\end{array}
\right. \;\;, 
\end{equation}
what produces the element of line 
\begin{equation}
ds^2=d\chi ^2+\sinh ^2\chi \left( \cos 2\theta +\cos 2\phi \right) \left( 
\frac{d\theta ^2}{\cos 2\theta }+\frac{d\phi ^2}{\cos 2\phi }\right) \;\;, 
\end{equation}
where yet $-\pi /4\leq \theta ,\phi \leq \pi /4$.

\subsubsection{Coordinates related to the hyperbolic parametrization}

There are two immediate distinct ways of generalizing the 2-dimensional set
of hyperbolic coordinates $\left( \rho ,\omega \right) $. The first one is 
\begin{equation}
\left\{ 
\begin{array}{lcl}
x^1 & = & \sinh \rho \cos \varphi \\ 
x^2 & = & \sinh \rho \sin \varphi \\ 
x^3 & = & \cosh \rho \sinh \omega \\ 
x^4 & = & \cosh \rho \cosh \omega 
\end{array}
\right. \;\;, 
\end{equation}
with element of line 
\begin{equation}
ds^2=d\rho ^2+\cosh ^2\rho d\omega ^2+\sinh ^2\rho d\varphi ^2\;\;. 
\end{equation}
The second one is 
\begin{equation}
\left\{ 
\begin{array}{lcl}
x^1 & = & \sinh \rho \\ 
x^2 & = & \cosh \rho \sinh \gamma \\ 
x^3 & = & \cosh \rho \cosh \gamma \sinh \omega \\ 
x^4 & = & \cosh \rho \cosh \gamma \cosh \omega 
\end{array}
\right. \;\;, 
\end{equation}
with element of line 
\begin{equation}
ds^2=d\rho ^2+\cosh ^2\rho \left( d\gamma ^2+\cosh ^2\gamma d\omega
^2\right) \;\;. 
\end{equation}
By noticing that this second element of line has a part conformally
equivalent to an hyperbolic 2-dimensional space one can write down other
three parametrizations by simply changing this part by any one of the set of
coordinates described in the previous section.

So, one can write 
\begin{equation}
\left\{ 
\begin{array}{lcl}
x^1 & = & \cosh \rho \sinh \varsigma \cos \varphi \\ 
x^2 & = & \cosh \rho \sinh \varsigma \sin \varphi \\ 
x^3 & = & \sinh \rho \\ 
x^4 & = & \cosh \rho \cosh \varsigma 
\end{array}
\right. \;\;, 
\end{equation}
obtaining then the element of line 
\begin{equation}
ds^2=d\rho ^2+\cosh ^2\rho \left( d\varsigma ^2+\sinh ^2\varsigma d\varphi
^2\right) \;\;. 
\end{equation}

In the same way one can have 
\begin{equation}
\left\{ 
\begin{array}{lcl}
x^1 & = & e^{-\sigma }\mu \cosh \rho \\ 
x^2 & = & \sinh \rho \\ 
x^3 & = & \left( \sinh \sigma +\frac 12e^{-\sigma }\mu ^2\right) \cosh \rho
\\ 
x^4 & = & \left( \cosh \sigma +\frac 12e^{-\sigma }\mu ^2\right) \cosh \rho 
\end{array}
\right. \;\;, 
\end{equation}
a parametrization with element of line 
\begin{equation}
ds^2=d\rho ^2+\cosh ^2\rho \left( d\sigma ^2+e^{-2\sigma }d\mu ^2\right)
\;\;. 
\end{equation}

Finally, one can write 
\begin{equation}
\left\{ 
\begin{array}{lcl}
x^1 & = & \sqrt{2}\cosh \rho \sinh u\cosh v \\ x^2 & = & \sqrt{2}\cosh \rho
\sinh v\cosh u \\ x^3 & = & \sinh \rho \\ 
x^4 & = & \cosh \rho \sqrt{\cosh 2u\cosh 2v} 
\end{array}
\right. \;\;, 
\end{equation}
obtaining then the element of line 
\begin{equation}
ds^2=d\rho ^2+\cosh ^2\rho \left( \cosh 2u+\cosh 2v\right) \left( \frac{du^2%
}{\cosh 2u}+\frac{dv^2}{\cosh 2v}\right) \;\;. 
\end{equation}

\subsubsection{Coordinates related to the exponential parametrization}

The exponential 2-dimensional coordinates $\left( \sigma ,\mu \right) $ can
be generalized for the space $H^3$ by the parametric equations 
\begin{equation}
\left\{ 
\begin{array}{lcl}
x^1 & = & e^{-\sigma }\mu \\ 
x^2 & = & e^{-\sigma }\nu \\ 
x^3 & = & \sinh \sigma +e^{-\sigma }\left( \mu ^2+\nu ^2\right) /2 \\ 
x^4 & = & \cosh \sigma +e^{-\sigma }\left( \mu ^2+\nu ^2\right) /2 
\end{array}
\right. \;\;, 
\end{equation}
with the subsequent element of line 
\begin{equation}
\label{exp3d}ds^2=d\sigma ^2+e^{-2\sigma }\left( d\mu ^2+d\nu ^2\right)
\;\;. 
\end{equation}
Since the last part of this element of line is conformal to a flat
2-dimensional space\footnote{%
In the same way of the 2-dimensional space $H^2$ here the substitution $%
z=e^{\sigma }$, with $0< z\leq\infty $, gives $ds^2=\left( d\mu ^2+d\nu
^2+dz^2\right) /z^2$, the element of line which represents the upper
half-space model of the space $H^3$ \cite{Thurston}.}, one can change it for
any of the parametrizations of the flat space and still obtains a coordinate
system where the Helmholtz equation is separable.

The first of these sets of coordinates is 
\begin{equation}
\left\{ 
\begin{array}{lcl}
x^1 & = & e^{-\sigma }\rho\cos\varphi \\ 
x^2 & = & e^{-\sigma }\rho\sin\varphi \\ 
x^3 & = & \sinh\sigma + 
\frac{1}{2}e^{-\sigma }\rho ^2 \\ x^4 & = & \cosh\sigma +\frac{1}{2}%
e^{-\sigma }\rho ^2 
\end{array}
\right.\;\;, 
\end{equation}
with the subsequent element of line 
\begin{equation}
ds^2=d\sigma ^2+e^{ -2\sigma }\left( d\rho ^2+\rho ^2 d\varphi
^2\right)\;\;. 
\end{equation}

Another one is 
\begin{equation}
\left\{ 
\begin{array}{lcl}
x^1 & = & ae^{-\sigma }\cosh u\cos v \\ 
x^2 & = & ae^{-\sigma }\sinh u\sin v \\ 
x^3 & = & \sinh\sigma + 
\frac{1}{2}a^2e^{-\sigma }\left(\cosh ^2 u-\sin ^2 v\right) \\ x^4 & = & 
\cosh\sigma +\frac{1}{2}a^2e^{-\sigma }\left(\cosh ^2 u-\sin ^2 v\right) 
\end{array}
\right.\;\;, 
\end{equation}
with element of line 
\begin{equation}
ds^2=d\sigma ^2+a^2e^{ -2\sigma }\left(\sinh ^2u+\sin^2v\right)\left( du
^2+dv ^2\right)\;\;. 
\end{equation}

A third system of coordinates following this line\footnote{%
Also following this line there is an interesting example of a system of
coordinates where the Helmholtz equation {\it is not} separable: 
\[
\left\{ 
\begin{array}{lcl}
x^1 & = & ae^{-\sigma }\sinh\eta \left(\cosh\eta-\cos\xi\right) ^{-1} \\ 
x^2 & = & ae^{-\sigma }\sin\xi \left(\cosh\eta-\cos\xi\right) ^{-1} \\ 
x^3 & = & \sinh\sigma +a^2e^{-\sigma }\left(\cosh\eta +\cos\xi \right)
\left(\cosh\eta -\cos\xi \right) ^{-1} \\ 
x^4 & = & \cosh\sigma +a^2e^{-\sigma }\left(\cosh\eta +\cos\xi \right)
\left(\cosh\eta -\cos\xi \right) ^{-1} 
\end{array}
\right.\;\;, 
\]
with element of line 
\[
ds^2=d\sigma ^2+a^2e^{ -2\sigma }\left(\cosh\eta-\cos\xi\right) ^{-2}\left(
d\eta ^2+d\xi ^2\right) \;\;. 
\]
} is 
\begin{equation}
\left\{ 
\begin{array}{lcl}
x^1 & = & e^{-\sigma }\xi\eta \\ 
x^2 & = & \frac{1}{2}e^{-\sigma }\left(\eta ^2-\xi ^2\right) \\ x^3 & = & 
\sinh\sigma + 
\frac{1}{8}e^{-\sigma }\left(\eta ^2+\xi ^2\right) ^2 \\ x^4 & = & 
\cosh\sigma +\frac{1}{8}e^{-\sigma }\left(\eta ^2+\xi ^2\right) ^2 
\end{array}
\right.\;\;, 
\end{equation}
with the subsequent element of line 
\begin{equation}
ds^2=d\sigma ^2+e^{ -2\sigma }\left(\eta ^2+\xi ^2\right)\left( d\eta
^2+d\xi ^2\right)\;\;. 
\end{equation}

\subsubsection{Coordinates related to the symmetric parametrization}

It is not easy to build a symmetric parametrization for the space $H^3$.
However, it is very simple to find one almost symmetric generalization of
the last coordinate system presented in the section \ref{2d}. This
generalization is 
\begin{equation}
\left\{ 
\begin{array}{lcl}
x^1 & = & \sqrt{2}\cosh u\sinh v \\ x^2 & = & \sqrt{2}\cosh v\sinh u \\ x^3
& = & \sqrt{\cosh 2u\cosh 2v}\sinh \chi \\ x^4 & = & \sqrt{\cosh 2u\cosh 2v}%
\cosh \chi 
\end{array}
\right. \;\;, 
\end{equation}
with element of line 
\begin{equation}
ds^2=\cosh 2u\cosh 2vd\chi ^2+\left( \cosh 2u+\cosh 2v\right) \left( \frac{%
du^2}{\cosh 2u}+\frac{dv^2}{\cosh 2v}\right) \;\;. 
\end{equation}

Table \ref{t2} presents a summary of all coordinate systems shown in this
section for the space $H^3$. The names used there for each system are
arbitrary and will be used in the next subsection.

\begin{table}[tbp]
\begin{center}
\begin{tabular}{|l|c|l|}
\hline
\multicolumn{3}{|c|}{{\bf Coordinate systems for the hyperbolic 3-D space}} \\ \hline
{\it system} & {\it coords.} & {\it element of line} \\ \hline
spherical polar & $\chi ,\theta ,\varphi $ & $d\chi^2+\sinh ^2\chi\left(d\theta ^2+\sin ^2\theta d\varphi ^2\right)$ \\ \hline
semi-spherical polar & $\chi ,\theta ,\phi $ & $d\chi^2$ \\
& & $+\sinh ^2\chi\left( 1+\cos 2\phi\sec 2\theta\right) d\theta ^2 $ \\
& & $+\sinh ^2\chi\left( 1+\cos 2\theta\sec 2\phi \right) d\phi ^2$ \\ \hline
hyperbolic & $\rho ,\omega ,\varphi $ & $d\rho ^2+\cosh ^2\rho d\omega ^2+\sinh ^2\rho d\varphi ^2$ \\ \hline
bi-hyperbolic & $\rho ,\omega ,\gamma $ & $d\rho ^2+\cosh ^2\rho \left( d\gamma ^2+\cosh ^2\gamma d\omega ^2\right)$ \\ \hline
polar hyperbolic & $\rho ,\varsigma ,\varphi $ & $d\rho ^2+\cosh ^2\rho \left( d\varsigma ^2+\sinh ^2\varsigma
d\varphi ^2\right)$ \\ \hline
exponential hyperbolic & $\rho ,\sigma ,\mu $ & $d\rho ^2+\cosh ^2\rho \left( d\sigma ^2+e^{-2\sigma }d\mu ^2\right)$ \\ \hline
symmetric hyperbolic & $\rho ,u,v $ & $d\rho ^2$ \\
& & $+\cosh ^2\rho \left( 1+\cosh 2u/\cosh 2v\right) dv^2 $ \\
& & $+\cosh ^2\rho \left( 1+\cosh 2v/\cosh 2u\right) du^2 $ \\ \hline
exponential & $\sigma ,\mu ,\nu $ & $d\sigma ^2+e^{ -2\sigma }\left( d\mu ^2+d\nu ^2\right)$ \\ \hline
polar exponential & $\sigma ,\rho ,\varphi $ & $d\sigma ^2+e^{ -2\sigma }\left( d\rho ^2+\rho ^2 d\varphi ^2\right)$ \\ \hline
elliptic exponential & $\sigma ,u,v $ & $d\sigma ^2$ \\
& & $+a^2e^{ -2\sigma }\left(\sinh ^2u+\sin^2v\right)\left( du ^2+dv ^2\right)$ \\ \hline
parabolic exponential & $\sigma ,\xi ,\eta $ & $d\sigma ^2+e^{ -2\sigma }\left(\eta ^2+\xi ^2\right)\left( d\eta ^2+d\xi ^2\right)$ \\ \hline
symmetric & $\chi ,u,v$ & $\cosh 2u\cosh 2v d\chi ^2$ \\
& & $+\left( 1+\cosh 2u/\cosh 2v\right) dv^2 $ \\
& & $+\left( 1+\cosh 2v/\cosh 2u\right) du^2 $ \\ \hline
bipolar exponential & $\sigma ,\xi ,\eta $ & $d\sigma ^2$ \\
& & $+a^2e^{ -2\sigma }\left(\cosh\eta-\cos\xi\right) ^{-2}\left( d\eta ^2+d\xi ^2\right)$ \\ \hline
\end{tabular}
\end{center}
\caption{\label{t2} Summary of systems for space $H^3$. Notice that the last one of these systems, named bipolar exponential, does not allow complete separation of the Helmholtz equation.}
\end{table}

\subsection{Helmholtz equation}

All the twelve coordinate systems presented in the previous subsection allow
to separate the Helmholtz equation, still given by the formula in equation (%
\ref{Helmholtz}), in a system of three differential equations, and these
equations are shown here. However, not all the differential equations
obtained are easy to solve analytically, and, therefore, in this paper some
of the solutions are not presented. In fact, one such case, concerning the
2-dimensional symmetric coordinates, already appeared at the end of section 
\ref{2d}.

\subsubsection{Polar coordinates}

The polar coordinates presented in equations (\ref{trintaedois}) and (\ref
{trintaetres}) produce the most known solutions of Helmholtz equation for
the space $H^3$. Such solutions are obtained by use of the tentative
function $\Psi \left( \chi ,\theta ,\varphi \right) =X\left( \chi \right)
Y\left( \theta ,\varphi \right) $, yielding then the system of equations 
\begin{equation}
\label{tdcp4}\frac{\partial ^2Y}{\partial \theta ^2}+\frac 1{\tan \theta }%
\frac{\partial Y}{\partial \theta }+\frac 1{\sin ^2\theta }\frac{\partial ^2Y%
}{\partial \varphi ^2}=-\ell \left( \ell +1\right) Y 
\end{equation}
and 
\begin{equation}
\label{tdcp5}\frac{d^2X}{d\chi ^2}+\frac 2{\tanh \chi }\frac{dX}{d\chi }%
+\left[ k^2-\frac{\ell \left( \ell +1\right) }{\sinh ^2\chi }\right]
X=0\,\,. 
\end{equation}
While the first, angular, equation admits as solutions the functions known
as spherical harmonics, $Y_{\ell m}\left( \theta ,\varphi \right) $, the
last equation, radial, also known as hyperspherical Bessel equation \cite
{Kosowsky}, produces 
\begin{equation}
\label{tdcp6a}
X =
\left( 1-\cosh ^2\chi\right)
^{-1/4}\left[ a_5P_{-\frac 12+\sqrt{1-k^2}}^{\ell +\frac 12}%
\left( \cosh\chi\right) +a_6Q_{-\frac 12+\sqrt{1-k^2}}^{\ell +\frac 12}\left( \cosh\chi\right) \right] \;\;,
\end{equation}
or
\begin{equation}
\label{tdcp6b}
X =\left( 1-\coth ^2\chi\right)
^{1/2}\left[ a_5^{\prime }P_{\ell }^{\sqrt{1-k^2}}\left( \coth\chi\right) +a_6^{\prime }Q_{\ell }^{\sqrt{1-k^2}}\left( \coth\chi\right) \right] \,\,.
\end{equation}
Another possible solution for the radial equation is found for $\kappa
_1=\cosh \chi $ by use of the `Leibniz's theorem', 
\begin{equation}
\label{tdcp8}\frac{d^m}{dx^m}\left[ f\left( x\right) g\left( x\right)
\right] =\sum_{n=0}^m\left( 
\begin{array}{c}
m \\ 
n 
\end{array}
\right) \frac{d^{m-n}}{dx^{m-n}}f\left( x\right) \frac{d^n}{dx^n}g\left(
x\right) 
\end{equation}
in the equation 
\begin{equation}
\label{tdcp9}\left( \kappa _1^2-1\right) \frac{d^2Z}{d\kappa _1^2}+\kappa _1%
\frac{dZ}{d\kappa _1}-\left( 1-k^2\right) Z=0 
\end{equation}
where 
\begin{equation}
\label{tdcp10}Z=\frac 12\left[ \left( \kappa _1+\sqrt{\kappa _1^2-1}\right)
^{1-k^2}+\left( \kappa _1-\sqrt{\kappa _1^2-1}\right) ^{1-k^2}\right] \,\,. 
\end{equation}
The coefficients obtained for each term must be compared to the ones in the
equation obtained from (\ref{tdcp5}) by the tentative function 
\begin{equation}
\label{tdcp11}X=\left( 1-\kappa _1^2\right) ^aY\,\,. 
\end{equation}
Such procedure shows that \cite{Cornish2,Cornish3,Hu} 
\begin{equation}
\label{tdcp12}X=a_{\sqrt{1-k^2}}^\ell \sinh ^\ell \chi \frac{d^{\ell
+1}\cosh \sqrt{\left( 1-k^2\right) \chi ^2}}{d\left(\cosh\chi\right) ^{\ell +1} } 
\end{equation}
where $a_{\sqrt{1-k^2}}^\ell $ is a constant. Just for completeness, it must
be cited that a solution of equation (\ref{tdcp5}) in terms of Gegenbauer
functions is also possible.

The coordinate system based in the parametrization of a semi-sphere produces
three separate differential equations by use of the function $\Psi \left(
\chi ,\theta ,\phi \right) =X\left( \chi \right) \Theta \left( \theta
\right) \Phi \left( \phi \right) $. These equations are equation (\ref{tdcp5}%
), and the pair 
\begin{equation}
\frac{d^2\Theta }{d\theta ^2}-\tan 2\theta \frac{d\Theta }{d\theta }+\left[
\ell \left( \ell +1\right) +\frac{\lambda ^2}{\cos 2\theta }\right] \Theta
=0 
\end{equation}
and 
\begin{equation}
\frac{d^2\Phi }{d\phi ^2}-\tan 2\phi \frac{d\Phi }{d\phi }+\left[ \ell
\left( \ell +1\right) -\frac{\lambda ^2}{\cos 2\phi }\right] \Phi =0\;. 
\end{equation}
Unfortunately, it is not easy to find analytical solutions for these two last
equations, except for the case $\lambda =0$, when the solutions are%
\begin{equation}
\Theta =\left( \cos \theta \right) ^{^{\frac 14}}\left[ aP_{\frac 14\left(
2\ell -1\right) }^{\frac 14}\left( \sin \theta \right) +bQ_{\frac 14\left(
2\ell -1\right) }^{\frac 14}\left( \sin \theta \right) \right] \;\;, 
\end{equation}
and 
\begin{equation}
\Phi =\left( \cos \phi \right) ^{^{\frac 14}}\left[
a^{\prime }P_{\frac 14\left( 2\ell -1\right) }^{\frac 14}\left( \sin \phi
\right) +b^{\prime }Q_{\frac 14\left( 2\ell -1\right) }^{\frac 14}\left(
\sin \phi \right) \right] \;. 
\end{equation}

\subsubsection{Group of the hyperbolic coordinates}

Using a solution of the form 
\begin{equation}
\label{tdf}\Psi \left( \rho ,\omega ,\varphi \right) =R\left( \rho \right)
\Omega \left( \omega \right) \Phi \left( \varphi \right) 
\end{equation}
one obtains, for the first set of coordinates derivated from the
2-dimensional hyperbolic coordinates, the differential equations 
\begin{equation}
\label{tdg}\frac{d^2\Omega }{d\omega ^2}=-\lambda ^2\Omega \;,\;\;\frac{%
d^2\Phi }{d\varphi ^2}=-\ell ^2\Phi \;, 
\end{equation}
two equations whose solutions are given by linear combinations of a sine and
a cosine, and 
\begin{equation}
\label{tdi}\frac{d^2R}{d\rho ^2}+\frac{\sinh ^2\rho +\cosh ^2\rho }{\sinh
\rho \cosh \rho }\frac{dR}{d\rho }-\left[ \frac{\lambda ^2}{\cosh ^2\rho }+%
\frac{\ell ^2}{\sinh ^2\rho }-k^2\right] R=0\,\,. 
\end{equation}
This last differential equation can be solved by the transformation of
coordinates 
\begin{equation}
\label{tdj}\tau =\cosh ^{-2}\rho \Rightarrow \tau ^{-2}d\tau =-2\sinh \rho
\cosh \rho d\rho 
\end{equation}
which yields solutions of the kind 
\begin{equation}
\label{tdl}R=\left( 1-\tau \right) ^p\tau ^qS 
\end{equation}
with 
\begin{equation}
\label{tdm}p=\frac \ell 2\;,\;\;q=\frac 12\left[ 1+\sqrt{1-k^2}\right] 
\end{equation}
and 
\begin{eqnarray}
\label{tdo}S&=&aF\left[ p+q+i\frac{\lambda }{2},p+q-i\frac{\lambda }{2}
;2q;\tau \right] \nonumber\\
&+&b\tau ^{1-2q}F\left[ p+\left( 1-q\right) +i\frac{
\lambda }{2},p+\left( 1-q\right) -i\frac{\lambda }{2};2\left( 1-q\right)
;\tau \right] \;\;,
\end{eqnarray}
where $F\left( \alpha ,\beta ;\gamma ;z\right) $ is an hypergeometric
function. Another equivalent solution, linked to this one by relations
between hypergeometric functions, is obtained from the substitution $\tau
^{\prime }=\tanh ^2\rho $.

For the second set of this group, named bi-hyperbolic, a function of the
form 
\begin{equation}
\label{tdt}\Psi =\Gamma \left( \gamma \right) R\left( \rho \right) \Omega
\left( \omega \right) 
\end{equation}
also divides the problem in three differential equations, 
\begin{equation}
\label{tdg1}\frac{d^2\Omega }{d\omega ^2}=-\ell ^2\Omega \;\;, 
\end{equation}
\begin{equation}
\label{tdh1}\frac{d^2\Gamma }{d\gamma ^2}+\tanh \gamma \frac{d\Gamma }{%
d\gamma }-\frac{\ell ^2}{\cosh ^2\gamma }\Gamma =-\lambda ^2\Gamma 
\end{equation}
and 
\begin{equation}
\label{tdi1}\frac{d^2R}{d\rho ^2}+2\tanh \rho \frac{dR}{d\rho }-\left[ \frac{%
\lambda ^2}{\cosh ^2\rho }-k^2\right] R=0\;. 
\end{equation}
The two first equations are equivalent to the equations obtained from (\ref
{wrau}), and therefore have solutions equivalent to (\ref{w}) and (\ref{rau}%
), while the last one produces 
\begin{equation}
\label{tdu}R =
\left(\cosh\rho\right) ^{-\frac12 }\left[aP_{-\frac 12 +\sqrt{1-k^2}}^{\sqrt{\frac 14-\lambda ^2}}\left( i\sinh\rho\right) +bQ_{-\frac 12 +\sqrt{1-k^2}}^{\sqrt{\frac 14-\lambda ^2}}\left( i\sinh\rho\right) \right] 
\end{equation}
or
\begin{equation}
\label{tdu2} 
R=\sqrt{\cosh\rho }\left[ a^{\prime }P_{-\frac 12 +\sqrt{\frac 14-\lambda ^2}}^{ 
\sqrt{1-k^2}}\left( \tanh\rho\right) +b^{\prime }Q_{-\frac 12 +\sqrt{\frac 14-\lambda ^2}}^{\sqrt{1-k^2}}\left( \tanh\rho\right) \right]
\;.
\end{equation}

For the third set of coordinates, $\left( \rho ,\varsigma ,\varphi \right) $%
, named polar hyperbolic, one can use a tentative function of the type $\Psi
\left( \sigma ,\varsigma ,\varphi \right) =R\left( \rho \right) \Xi \left(
\varsigma \right) \Phi \left( \varphi \right) $, and thus obtain equations
similar to (\ref{tdg1}) and (\ref{tdi1}) for the functions $\Phi \left(
\varphi \right) $ and $R\left( \rho \right) $, respectively, and the
equation 
\begin{equation}
\frac{d^2\Xi }{d\varsigma ^2}+\coth \varsigma \frac{d\Xi }{d\varsigma }%
-\left( \frac{\ell ^2}{\sinh ^2\varsigma }-\lambda ^2\right) \Xi =0\;, 
\end{equation}
whose solution is equivalent to the one given in (\ref{vinteedoisb}).

The fourth set of coordinates of this group, named exponential hyperbolic,
allows the use of the function $\Psi \left( \sigma ,\mu ,\rho \right)
=\Sigma \left( \sigma \right) M\left( \mu \right) R\left( \rho \right) $ to
produce, beyond eq. (\ref{tdi1}), the equations 
\begin{equation}
\frac{d^2M}{d\mu ^2}+\ell ^2M=0\;, 
\end{equation}
equivalent to (\ref{tdg1}), and 
\begin{equation}
\frac{d^2\Sigma }{d\sigma ^2}-\frac{d\Sigma }{d\sigma }-\left( e^{2\sigma
}\ell ^2-\lambda ^2\right) \Sigma =0\;, 
\end{equation}
whose solution appears in (\ref{vinteesete}).

Finally, the function $\Psi \left( \rho ,u,v\right) =R\left( \rho \right)
U\left( u\right) V\left( v\right) $ produces for the symmetric hyperbolic
coordinates the equations (\ref{tdi1}) and the pair 
\begin{equation}
\frac{d^2U}{du^2}+\tanh 2u\frac{dU}{du}+\left( \lambda ^2-\frac{\ell ^2}{%
\cosh 2u}\right) U=0 
\end{equation}
and 
\begin{equation}
\frac{d^2V}{dv^2}+\tanh 2v\frac{dV}{dv}+\left( \lambda ^2+\frac{\ell ^2}{%
\cosh 2v}\right) V=0\;, 
\end{equation}
equivalent to (\ref{vinteenove}) e (\ref{trinta}).

\subsubsection{Group of the exponential coordinates}

The metric present in the element of line given by equation (\ref{exp3d}),
of the exponential coordinates, when in combination with the tentative
function $\Psi \left( \sigma ,\mu ,\nu \right) =\Sigma \left( \sigma \right)
M\left( \mu \right) N\left( \nu \right) $, produces from the Helmholtz
equation three equations, 
\begin{equation}
\frac{d^2M}{d\mu ^2}+\ell ^2M=0\;,\;\;\frac{d^2N}{d\nu ^2}+\left( \lambda
^2-\ell ^2\right) N=0\;, 
\end{equation}
solved in terms of sines and cosines, and 
\begin{equation}
\label{smodel}\frac{d^2\Sigma }{d\sigma ^2}-2\frac{d\Sigma }{d\sigma }%
-\left( e^{2\sigma }\lambda ^2-k^2\right) \Sigma =0\;, 
\end{equation}
whose solution is 
\begin{equation}
\label{shazam}\Sigma \left( \sigma \right) =e^\sigma \left[ bI_{\sqrt{1-k^2}%
}\left( \lambda e^\sigma \right) +cK_{\sqrt{1-k^2}}\left( \lambda e^\sigma
\right) \right] \;. 
\end{equation}

The parametrization with coordinates $\left( \sigma ,\rho ,\varphi \right) $%
, named here polar exponential, produces the equations 
\begin{equation}
\frac{d^2\Phi }{d\varphi ^2}+\ell ^2\Phi =0\;, 
\end{equation}
solved in terms of sines and cosines, 
\begin{equation}
\frac{d^2R}{d\rho ^2}+\frac 1\rho \frac{dR}{d\rho }+\left( \lambda ^2-\frac{%
\ell ^2}{\rho ^2}\right) R=0\;, 
\end{equation}
which is Bessel's equation for the variable $\lambda \rho $, and eq. (\ref
{smodel}).

With the coordinates present in the third parametrization derivated of the
2-dimensional exponential coordinates, named elliptic exponential, one can
built a tentative function $\Psi \left( \sigma ,\mu ,\nu \right) =\Sigma
\left( \sigma \right) M\left( \mu \right) N\left( \nu \right) $, and thus
obtain equation (\ref{smodel}) and the pair 
\begin{equation}
\frac{d^2U}{du^2}-\left( a^2\lambda ^2\sinh ^2u+\ell ^2\right) U=0 
\end{equation}
and 
\begin{equation}
\frac{d^2V}{dv^2}-\left( a^2\lambda ^2\sin ^2v-\ell ^2\right) V=0\;. 
\end{equation}
These two equations are Mathieu's modified equation and Mathieu's equation,
respectively, whose solutions are given in terms of Mathieu functions \cite
{GR}.

Finally, one can obtain for the fourth set of coordinates in this group,
named parabolic exponential, again equation (\ref{smodel}) and the pair 
\begin{equation}
\frac{d^2N}{d\eta }-\left( \lambda ^2\eta ^2+\ell ^2\right) N=0 
\end{equation}
and 
\begin{equation}
\frac{d^2\Xi }{d\xi ^2}-\left( \lambda ^2\xi ^2-\ell ^2\right) \Xi =0\;, 
\end{equation}
solved with the use of parabolic cylinder functions \cite{GR,Erdelyi}.

\subsubsection{Symmetric coordinates}

The almost symmetric coordinates $\left( \chi ,u,v\right) $ produce, by use
of the tentative function $\Psi \left( \chi ,u,v\right) =X\left( \chi
\right) U\left( u\right) V\left( v\right) $, the equation 
\begin{equation}
\frac{d^2X}{d\chi ^2}+\lambda ^2X=0\;,
\end{equation}
whose general solution is a linear combination of a sine and a cosine, and
the pair 
\begin{equation}
\frac{d^2U}{du^2}+2\tanh 2u\frac{dU}{du}+\left( k^2-\frac{\lambda ^2}{\cosh
^22u}-\frac{\ell ^2}{\cosh 2u}\right) U=0
\end{equation}
and 
\begin{equation}
\frac{d^2V}{dv^2}+2\tanh 2v\frac{dV}{dv}+\left( k^2-\frac{\lambda ^2}{\cosh
^22v}+\frac{\ell ^2}{\cosh 2v}\right) V=0\;.
\end{equation}
These two last equations have simple analytical solutions only when the
separation constant $\ell $ is null, a case in which one has%
\begin{equation}
U\left( u\right) =\left( \cosh 2u\right) ^{-\frac 12}\left[ aP_{-\frac
12\left( 1-i\lambda \right) }^{\frac 12
\sqrt{1-k^2}}\left( \tanh 2u\right) +bQ_{-\frac 12\left( 1-i\lambda \right)
}^{\frac 12\sqrt{1-k^2}}\left( \tanh 2u\right) \right] 
\end{equation}
and
\begin{equation} 
V\left( v\right) =\left( \cosh 2v\right) ^{-\frac 12}\left[ a^{^{\prime
}}P_{-\frac 12\left( 1-i\lambda \right) }^{\frac 12\sqrt{1-k^2}}\left( \tanh
2v\right) +b^{^{\prime }}Q_{-\frac 12\left( 1-i\lambda \right) }^{\frac 12%
\sqrt{1-k^2}}\left( \tanh 2v\right) \right] \;\;.
\end{equation}

\section{\label{end} Conclusion}

Specific boundary conditions imposed by a certain problem are, in general,
better dealt with by use of one specific general solution of the Helmholtz
equation obtained in some particular parametrization. This idea is easily
seen when one tries to find the adequate solutions to describe the movements
of waves inside a box: of all the 11 coordinate systems suitable for the
Euclidean 3-dimensional space only the Cartesian one is perfectly fit for
this job.

An interesting problem following this idea consists in finding the modes
allowed in a finite flat space known as flat $2$-torus, a non-trivial,
compact manifold represented by a rectangle with opposite sides identified.
Several equal copies of such manifold can be put side by side to fill
entirely the Euclidean plane $E^2$, in a regular tiling -- or tessellation
--, and therefore the adequate solutions of the Helmholtz equation in this
manifold must be such that they are periodic. It is not hard to see that the
adequate solutions are written as plane waves, {\it i.e.}, linear combinations of
sines and cosines of the rectangular Cartesian coordinates $x$ and $y$.

There are similar problems for the hyperbolic spaces. The representation of
the simplest compact 2-dimensional surface of negative curvature, the torus
of genus $2$, topologically equivalent to a double-doughnut, is built by the
identification of pairs of sides of a regular octagon \cite{Balazs}. In the Klein
projection the octagon appears with sides formed by segments of straight
lines, while in the Poincar\'e projection it appears as formed by arcs of
circles. Another examples of manifolds of this type can be seen in the
polyhedra that represent compact 3-dimensional hyperbolic spaces, drawn by
the software {\it SnapPea} \cite{Snappea} in the Klein projection.

For the hyperbolic case, however, differently from the Euclidean one, the
properties of compact --{\it i.e.}, finite -- spaces are far from being
completely known. Several studies with problems involving these spaces need
numerical integrations which demand considerable computational time -- see as examples refs. \cite{CBPF,Cornish+Turok,Inoue2}. The use
of adequate coordinates could facilitate such studies, as exemplified in a
recent paper where a problem of quantum cosmology, previously dealt with by
use of crude estimates, is reanalyzed with the use of an adequate coordinate
system, in a procedure that allowed also to calculate the
volume of non-trivial compact 3-dimensional hyperbolic spaces \cite{Costa}.
Following such line of research, this article, therefore,
simply presents a table of basic results that could be of some help for
those dealing with the study of hyperbolic spaces, in the hope of motivate
new developments in this field, particularly in the unveiling of properties
of hyperbolic compact spaces, which are of interest in cosmology.

\subsection*{Acknowledgments}
The author thanks the Brazilian agency FAPESP for the financial support (grant 00/13762-6).

\newpage 

\thispagestyle{empty}

\thispagestyle{empty}

\end{document}